# Giant peak of the Inverse Faraday effect in the band gap of magnetophotonic crystal

Mikhail A. Kozhaev,[1,2] Alexander I. Chernov,[1,2] Daria A. Sylgacheva,[1,3] Alexander N. Shaposhnikov,[4] Anatoly R. Prokopov,[4] Vladimir N. Berzhansky,[4] Anatoly K. Zvezdin,[1,2] Vladimir I. Belotelov[1,3,*]

[1] Russian Quantum Center, 45 Skolkovskoye shosse, Moscow, 121353, Russia

[2] Prokhorov General Physics Institute RAS, 38 Vavilov Street, Moscow 119991, Russia

[3] Faculty of Physics, Lomonosov Moscow State University, Leninskie Gory, Moscow 119991, Russia

[4] Vernadsky Crimean Federal University, 4 Vernadskogo Prospekt, Simferopol, 295007, Russia

Correspondence to belotelov@physics.msu.ru

**Optical impact on the spin system in a magnetically ordered medium provides a unique possibility for local manipulation of magnetization at subpicosecond time scales that is very promising for magnetic data processing and other magnonics applications. One of the mechanisms of the optical manipulation is related to the inverse Faraday effect (IFE). Usually the IFE is observed in crystals and magnetic films on a substrate. Here we demonstrate the IFE induced by fs-laser pulses in the magnetic film inside the magnetophotonic crystal. Spectral dependence of the IFE on the laser pulse wavelength in the band gap of the magnetophotonic crystal has a sharp peak leading to a significant enhancement of the IFE. This phenomenon is explained by strong confinement of the electromagnetic energy and angular momentum within the magnetic film. Calculated near field distribution of the IFE effective magnetic field indicates its subwavelength localization within 30 nm along the film thickness. These excited volumes can be shifted along the sample depth via e.g. changing frequency of the laser pulses. The obtained results open a way for the new applications in the areas of ultrafast spintronics and quantum information processing.**

**Introduction**

Optical control of the magnetization at ultrashort time scales is of prime interest[1-6] in context of the data processing and spintronic applications[7,8]. The inverse Faraday effect (IFE) provides a new way for efficient control of spins at GHz and THz rates[1]. Circularly polarized laser pulses can induce the IFE effective magnetic field of about $10^2$-$10^4$ Oe in a magnetic dielectric and excite the magnetization precession leading to magnetostatic spin waves[9-18]. By now, the IFE has been investigated only in a freestanding crystals and films on a substrate. In this case optical pulses influence on the magnetization almost uniformly along the whole depth of the magnetic film. In the spectral range where dispersion of the optical and magneto-optical properties of a magnetic sample is small the IFE depends on the laser pulse wavelength marginally.

At the same time, magnetic periodic structures like magneto-photonic and magneto-plasmonic crystals provide notable resonances in optical spectra resulting in significant enhancement of different magneto-optical effects[19-27]. The main reason of the enhancement is due to the excitation of the eigenmodes of the structure, which makes the light-matter interaction more efficient. For the magnetic structures containing conductive media the key role is played by surface plasmon-polaritons[19-24], while in magnetophotonic crystals (MPC) -- all-dielectric periodic multilayer magnetic structures -- the magneto-optical effects increase at the microcavity resonance inside the photonic band gap[25-27].

Optical control of magnetization using magnetic nanostructures has been attempted in GdFeCo and TbFeCo films covered with gold double-wire antennas and rectangular apertures in a gold film[28,29]. Subwavelength localization of the magnetic area influenced by light was achieved. Recent paper[30] describes the plasmon-induced demagnetization in nickel nanoparticles. The IFE in the samples covered by the nano-size apertures was theoretically investigated in the view of possible application for high-density all-optical magnetic recording[31].

Due to the large optical losses in metals the plasmonic resonances are quite broad, that does not allow obtaining large values of the IFE and rather leads to thermal effects. In this respect, all-dielectric structures are more advantageous since they provide optical resonances with high optical quality factors[25,26]. The IFE in all-dielectric structures has not yet been experimentally considered.

In this work, we investigate the influence of the optical confinement on the IFE in the all-dielectric MPCs. We excite the magnetization dynamics by the fs-laser pulses in the MPC to demonstrate that the IFE becomes very sensitive to the excitation laser wavelength and increases significantly at the microcavity resonance in the photonic band gap.

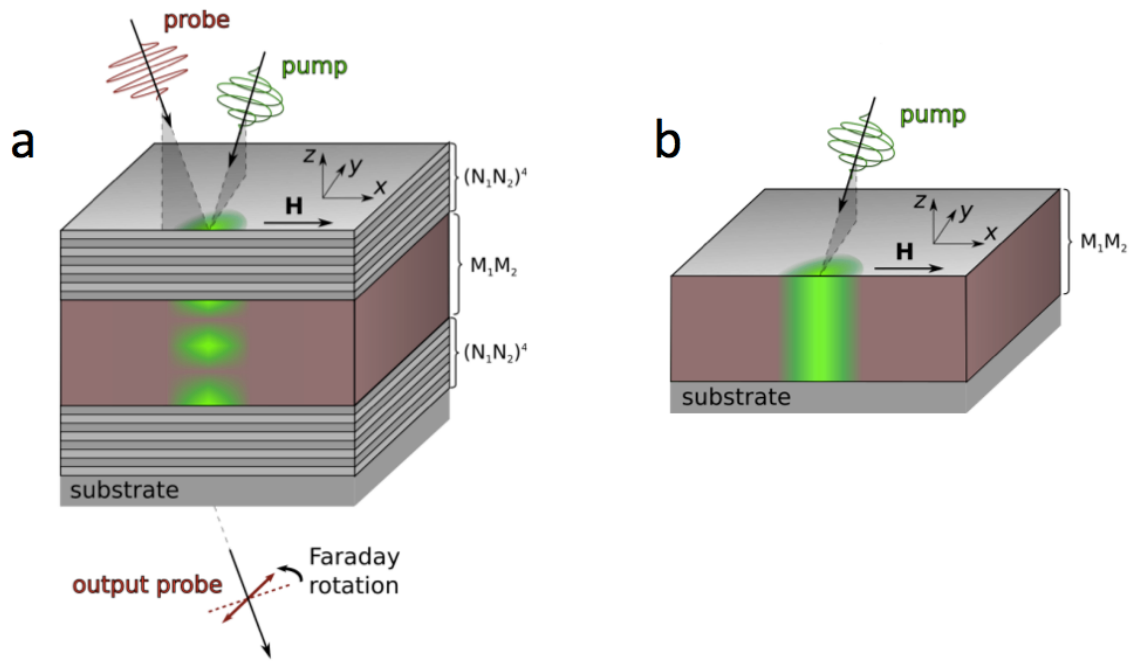

**Figure 1 | The 3D Inverse Faraday effect in a magnetophotonic crystal.** (a) Scheme of the experiment. The sample is a magnetophotonic crystal formed by the microcavity magnetic film (brown) sandwiched in between two nonmagnetic Bragg mirrors formed by several pairs of the dielectric layers N1 and N2 (gray). The circularly polarized pump excites the magnetic film and the linearly polarized probe is used to observe the magnetization dynamics at some time delay. (a,b) Calculated distributions of the optically generated effective magnetic field inside the magnetic layer of the MPC and inside a single magnetic film are shown on the front side of the samples by green color demonstrating the 3D and 2D localization of the IFE, respectively.

**Results**

For performing the experiments, we used the MPC with a bismuth iron-garnet magnetic film sandwiched between two nonmagnetic dielectric Bragg mirrors (Fig. 1a). Iron-garnet films doped with bismuth provide a large Faraday effect and have a relatively low optical absorption in the long-wavelengths visible spectral range [32,33]. Doping allows achieving high magneto-optical figure of merit in the MPC. The magnetic film has out-of-plane anisotropy and consists of the auxiliary layer (M1) and the main layer (M2). The M1-layer is needed to deposit the high quality M2-layer with large Bi concentration. For the 43° light incidence the cavity mode is excited in the photonic band gap center at $\lambda_0 = 642$ nm (Fig. 2a). It is accompanied by the 5-fold enhancement of the Faraday effect with respect to the single magnetic film. The optically excited spin dynamics in the MPC was investigated by the pump-probe experimental technique. The circularly polarized 150 fs pump-pulse was used to excite the magnetization precession due to the IFE: an effective magnetic field, $H_{IFE}$, appears in the sample during the pulse propagation.

The observed oscillations of the Faraday angle of the probe beam, $\Psi$, can be described by the harmonic function with decaying amplitude, $\Psi(t) = \Psi_m \exp(-t/\tau) \sin \omega t$, where $\Psi_m$ is an initial amplitude of $\Psi$, $\tau$ is a decay time of the precession, and $\omega$ is a precession frequency (Fig. 2b). Amplitude of the observed oscillations strongly depends on the pump wavelength that is varied near the cavity resonance. Full-width at half-maximum (FWHM) of the $\Psi_m(\lambda)$ resonance is $\Delta\lambda = 12$ nm, that is equal to the FWHM of the transmission peak.

It should be noted that the decay time is also sensitive to the pump wavelength: the longest decay appears at the resonance wavelength. It might be due to the fact that at the resonance wavelength the optical power concentration in the magnetic layer is maximal. As it was revealed previously, for the pumping of the iron-garnet films containing Gd ions at 600 – 700 nm the increase of the fluence makes $\tau$ larger[34].

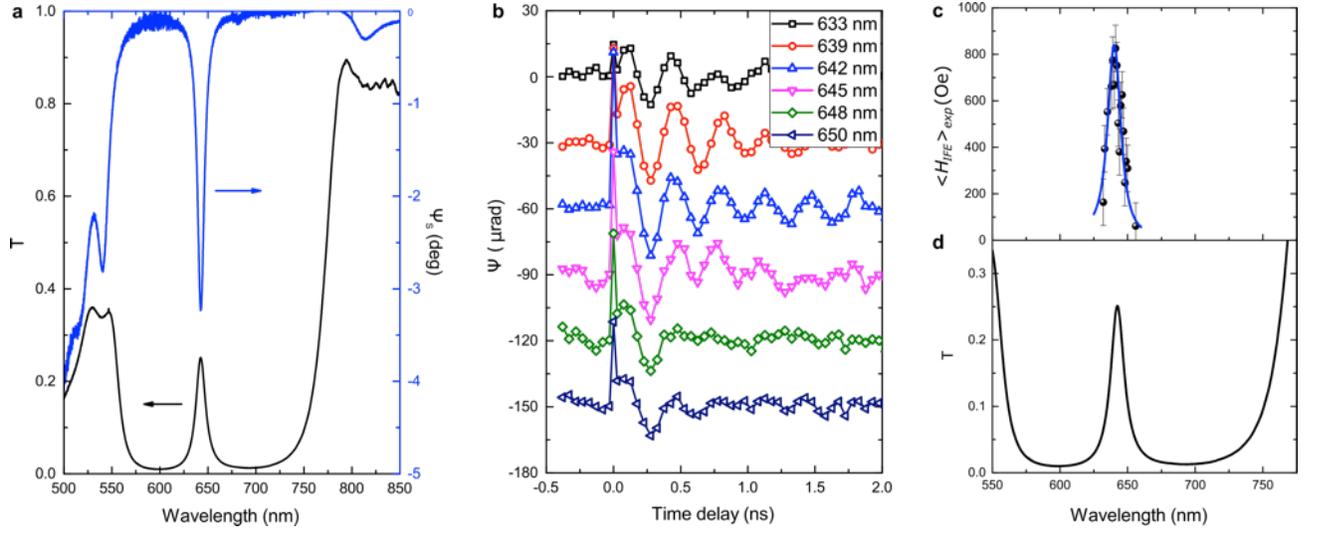

**Figure 2 | Laser pulse excited magnetization dynamics in MPC.** (a) Spectra of the optical transmittance (black curve) and Faraday rotation (blue curve) for the MPC fully magnetized out-of-plane. (b) Time-resolved change of the Faraday rotation indicating the magnetization precession at different excitation wavelengths around the cavity resonance within the photonic band gap. All curves have offsets for clarity of representation. (c) Averaged over the thickness of the magnetic film the normal component of the IFE magnetic field versus the pump wavelength: found from the experimental data, $\langle H_{IFE} \rangle_{exp}$ (black spheres), and calculated from the electromagnetic field distribution, $\langle H_{IFE} \rangle_{calc}$ (solid blue curve). External magnetic field is 890 Oe. The pump fluence is 0.66 mJ/cm². (d) Optical transmittance of the pump beam versus the pump wavelength. Angle of light incidence is 43°.

As duration of the laser pulses, $\Delta t$, is much smaller than the ferromagnetic resonance period, the magnetization dynamics can be considered as the decaying precession around the external magnetic field with the initial conditions determined by $H_{IFE}$. If direction of the magnetization vector is described by the angle $\theta$ between the magnetization and the film plane, then $\theta(t) = \theta_m e^{-\frac{t}{\tau}} \sin \omega t$, where $\theta_m$ is initial amplitude of $\theta$. The precession amplitude is mainly determined by the normal component of the IFE magnetic field, $H_{IFE}$:

$$\theta_m = \gamma (1 + H_a/H)^{-1/2} H_{IFE} \Delta t, \quad (1)$$

where $\gamma$ is a gyromagnetic ratio, $H_a = 4\pi M - 2K_U/M$, $M$ is a saturation magnetization and $K_U$ is an uniaxial anisotropy constant[34] (Supplementary material). Therefore, $H_{IFE}$ can be found from measuring the precession amplitude.

In the magnetic film inside the MPC the optical field distribution over the film thickness is not uniform, that causes some distribution of $\mathbf{H}_{IFE}(z)$. In accordance to Eq.(1) $\boldsymbol{\theta}_m(z) \sim \mathbf{H}_{IFE}(z)$. With time the distribution of $\boldsymbol{\theta}(z)$ becomes smoother since spins in the adjacent areas of the film interact by the dipole-dipole and exchange interactions.

Depth dependent magnetization precession $\boldsymbol{\theta}(z)$ is detected by the probe pulse. As the probe pulse also propagates through the MPC the distribution of its electric field, $\mathbf{E}_{pr}(z)$, in the magnetic film is not uniform as well. Consequently, the observed signal takes into account relative distribution of $\boldsymbol{\theta}_m(z)$ and $\mathbf{E}_{pr}(z)$. However, in order to estimate the out-of-plane component of the IFE magnetic field, $\langle H_{IFE}\rangle_{exp}$, from the experiment, one could assume that the probe beam measures averaged over the film thickness precession angle $\langle\theta\rangle$ and $\boldsymbol{\Psi}_m = \boldsymbol{\Psi}_s \cos\beta \langle\boldsymbol{\theta}_m\rangle$, where $\boldsymbol{\Psi}_s$ is the Faraday rotation of the film magnetically saturated out-of-plane and $\boldsymbol{\beta}$ is the refraction angle of the incident light inside the magnetic film. Calculations confirm that this assumption leads to rather small inaccuracy, but allows a notable simplification of the formulas. Consequently, average value of the IFE magnetic field, $\langle H_{IFE}\rangle_{exp}$, can be found from the experimental data as:

$$\langle H_{IFE}\rangle_{exp} = \frac{\sqrt{1+H_a/H}}{\gamma \cos\beta \Delta t} \frac{\Psi_m}{\Psi_s}. \quad (2)$$

The value of the IFE strongly depends on the pump wavelength (black circles in Fig. 2c) demonstrating a giant peak in the photonic band gap ($\lambda$=642nm). The spectral dependence is similar to the direct Faraday effect. At the resonance wavelength $\langle H_{IFE}\rangle_{exp}$ reaches 840 Oe.

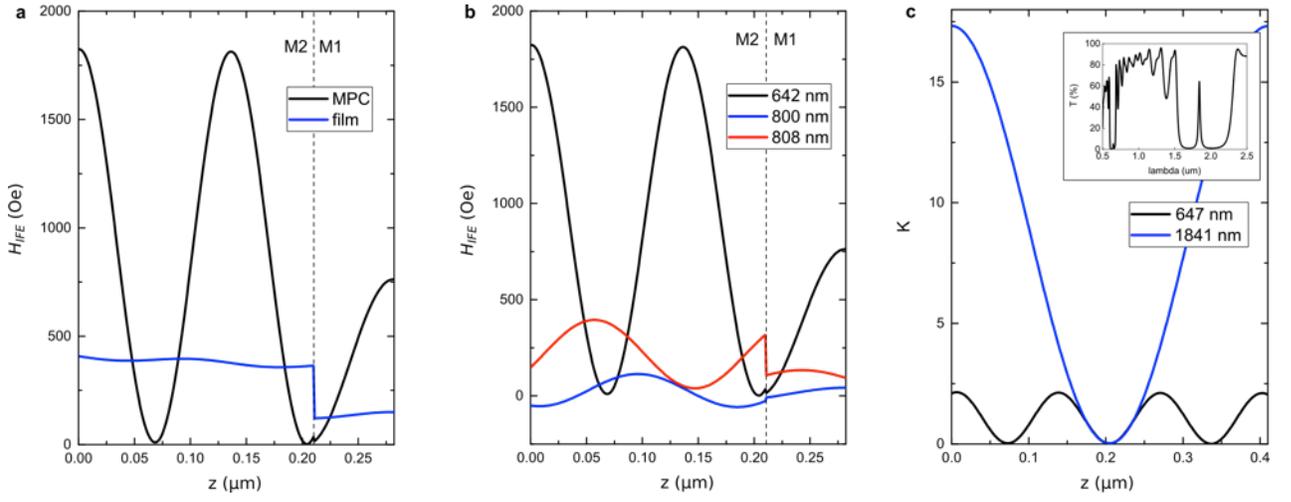

**Figure 3 | Calculated distribution of the IFE effective magnetic field in the experimentally studied MPC (a, b) and in the MPC with three time thicker layers (c).** (a) The IFE magnetic field in the magnetic film of the experimentally studied MPC at the cavity resonance (black curve) and in the same single magnetic film on the substrate (blue curve) at λ = 642 nm. (b) The IFE magnetic field in the magnetic film of the experimentally studied MPC at λ = 642 nm (black curve), 800 nm (blue curve), and 808 nm (red curve). In (a) and (b) two magnetic layers are shown: the main one (M1, 0 < z < 210 nm) and the auxiliary one (M2, 210 < z < 282 nm). (c) The IFE magnetic field enhancement factor, $K$, in the MPC with the main resonance at λ = 1841 nm as compared to the single magnetic film at the 1-st order (λ = 1841 nm, blue curve) and 2-nd order (λ = 647 nm, black curve) cavity resonances. The MPC in (c) has Bragg mirrors of four pairs of 228-nm-thick $TiO_2$ and 351-nm-thick $SiO_2$ layers, and the magnetic film of 410-nm-thick $Bi_{1.5}Gd_{1.5}Fe_{4.5}Al_{0.5}O_{12}$. Area of the magnetic layer is shown (0 < z < 410 nm). Inset: transmission spectrum of the MPC with the main resonance at λ = 1841 nm.

On the other hand, the IFE magnetic field is determined by the distribution of the pump electric field **E** inside the magnetic film[1]:

$$\mathbf{H}_{IFE} = -\frac{\varepsilon_0}{\mu_0}\frac{g}{M}\text{Im}([\mathbf{E}\times\mathbf{E}^*]), \quad (3)$$

where $\varepsilon_0$ and $\mu_0$ are free space permittivity and permeability, respectively, and $g$ is a medium gyration. To calculate the distribution of **E**(z) we used the transfer matrix method[35] (Supplementary material). At $\lambda$=642nm Eq.(3) gives $H_{IFE}(z)$ with maximum value of $(H_{IFEm})_{calc} = 1825$ Oe and the average of $\langle H_{IFE}\rangle_{calc} = 850$ Oe (Fig. 3a, black curve). The latter is in nice agreement with $\langle H_{IFE}\rangle_{exp}$ found from the experimental data via Eq. (2). Calculations for other pump wavelengths around the cavity resonance also follow the experimental results with good precision (solid blue curve in Fig. 2c). It confirms applicability of Eqs. (1)-(3) in our case.

At the cavity resonance ($\lambda$=642nm) the function of $H_{IFE}(z)$ in each magnetic layer takes harmonic form: $H_{IFE}(z) = H_{IFEm} \cos^2 \frac{2\pi\sqrt{\varepsilon}}{\lambda_0} z$, where $\varepsilon$ is the magnetic layer permittivity (black curve in Fig. 3a). The value of $H_{IFEm}$ exceeds by about 5 times the IFE field in the same single magnetic film on the substrate, $H_{IFE0}$ (blue curve in Fig. 3a). Distribution of $H_{IFE0}(z)$ is close to uniform.

The field $H_{IFE}(z)$ has three maxima within the magnetic film at $z = \zeta_i$. Two of them are at the interfaces with the Bragg mirrors: $\zeta_1 = 0$ with FWHM of $\frac{\lambda_0}{8\sqrt{\varepsilon_1}} = 33$ nm and $\zeta_3 = 282$ nm with FWHM of $\frac{\lambda_0}{8\sqrt{\varepsilon_2}} = 37$ nm. The intermediate maximum is at $\zeta_2 = \frac{\lambda_0}{2\sqrt{\varepsilon_1}} = 130$ nm and has FWHM of $\frac{\lambda_0}{4\sqrt{\varepsilon_1}} = 65$ nm. Therefore, the laser pulses excite the spin dynamics in the magnetic film locally in depth. Moreover, focusing of the pulses also adds the lateral locality. As a result, in the MPC we deal with the three-dimensional localization of the IFE in the cylindrical regions, which are 9 μm in diameter and only several tens of nanometers in height.

Furthermore, position of the IFE localization regions inside the magnetic film is tunable by variation of the pump wavelength or incidence angle. In particular, the IFE-regions are shifted by about 60 nm in depth, if the pump wavelength is changed from $\lambda$=642 nm to $\lambda$=808 nm, corresponding to the edge of the photonic band gap (Fig. 3b). Therefore, spins in the film layers that are not directly excited at $\lambda$ = 642 nm, become accessible at $\lambda$ = 808 nm, and vice versa. At some wavelength range (around $\lambda=800$ nm) the direction of the IFE magnetic field changes by the opposite one within only 100 nm depth (blue curve in Fig. 3b).

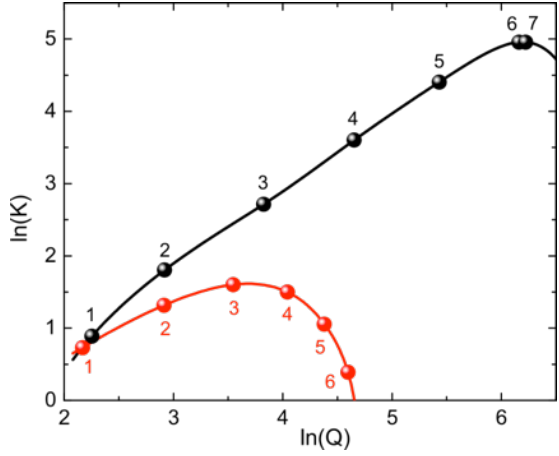

**Figure 4 | Enhancement factor of the inverse Faraday effect versus quality factor of the magnetophotonic crystals.** Two cases of different optical losses in the iron-garnet layers are considered: at $\lambda_0 = 642$ nm (red curve) and at $\lambda_0 = 1400$ nm (black curve). Numbers near the circles indicate number of the $N_1N_2$ dielectric layer pairs in the Bragg mirrors of the MPC.

If the MPC parameters are chosen to get the cavity modes of several orders in the transparency wavelength range of the magnetic film, then the IFE localization regions can be drastically changed by switching the pump between the cavity resonances (Fig. 3c). The difference in the enhancement coefficient $K = \frac{H_{IFEm}}{H_{IFE0}}$ at $\lambda = 647$ nm and $\lambda = 1841$ nm is due to much higher absorption of iron-garnets in the visible range. The maximum localization length is $\frac{\lambda_0}{8\sqrt{\varepsilon}}$ and, therefore, it becomes smaller for the higher-order resonances. However, its further decrease is limited by absorption at short-wavelengths. For iron-garnets the acceptable absorption range starts at $\lambda > 600$ nm which implies best localization of 35 nm.

Variation of the probe wavelength would also give additional advantage as it modifies distribution of the probing electromagnetic field and therefore provides sensitivity to the magnetization dynamics at different parts inside the magnetic film.

Resonance increase and localization of the IFE in the small regions inside the magnetic film at the wavelength of the cavity mode points out that the optical confinement is crucial for these phenomena. It can also be considered in terms of the increase of the optical density states in the cavity with respect to the free space in accordance to the Purcell effect.

The optical confinement is characterized by the quality factor, $Q$, defined by the ratio of the cavity resonance wavelength, $\lambda_0$, to the resonance full-width at half maximum, $\Delta\lambda$: $Q = \lambda_0/\Delta\lambda$. For the definite materials of the MPC $Q$ is determined by the number of layers in the Bragg mirrors.

For the experimentally studied MPC the absorption of the magnetic film makes the dependence of the enhancement coefficient $K(Q)$ non-monotonic (Fig. 4). At the level of absorption of the experimentally studied magnetic layers the optimal $Q$ corresponds to three pairs of $TiO_2/SiO_2$ layers in the MPC Bragg mirrors. Consequently, the investigated MPC sample is close to the optimal one. However, at the resonance wavelength of $\lambda_0 = 1400$ nm the absorption of the magnetic film is more than 550 times smaller which leads to the linear dependence of $K(Q)$ for up to the 6 pairs of the $TiO_2/SiO_2$ layers ($Q = 470$) and promises maximum magnitude of $K = 110$ (Fig. 4, black curve).

## Discussion

We have demonstrated that magnetophotonic crystals provide a giant peak in the dependence of the inverse Faraday effect on the pump wavelength and a significant enhancement of its value. In particular, the experimentally studied MPC gives a 5-time increase of the IFE with respect to a single film.

The enhancement factor is related to the quality factor of the microcavity and for this reason it is more advantageous to use dielectric Bragg mirrors rather than the metallic ones and operate in the transparency wavelength range of the magnetic film.

Moreover, calculation of the optical field distribution in the MPC reveals a subwavelength localization of the IFE along the film thickness down to tens of nanometers. Distribution of the IFE magnetic field in the magnetic film can be varied with the pump wavelength or the incidence angle of the fs-laser pulses. This feature can be used to access specific spins in the magnetophotonic crystal. Here we have found the IFE-region shift by about 60 nm for the pump wavelength change from 642 nm to 808 nm.

Optical confinement might allow excitation of spin wave resonances in the magnetic system, i.e. magnetic modes, which are characterized by non-uniform distribution over the film thickness[36]. Further development of our concept of the IFE in structures with optical confinement can be performed by the implementation of different types of plasmonic or all-dielectric nanoantennas to concentrate the optical fields and inverse Faraday effect within even smaller volumes with a thickness of about ten nanometers. Therefore, the results of this study pave a way for the new applications in such areas as spintronics, holographic memory and quantum information processing.

## Methods

**Magnetophotonic crystal sample.** The MPC structure consists of the magnetic film and two non-magnetic dielectric Bragg mirrors. The iron-garnet magnetic film was deposited in the argon-oxygen mixture by reactive ion beam sputtering of two targets with different compositions[28]. It includes two layers M1 and M2: the auxiliary 72-nm-thick $Bi_{1.0}Y_{0.5}Gd_{1.5}Fe_{4.2}Al_{0.8}O_{12}$ layer (the saturation magnetization is $4\pi M_s = 330$ **Gs**, and the uniaxial magnetic anisotropy constant is $K_U = -0.88 \cdot 10^3$ **erg/cm³**) and the main 210-nm-thick $Bi_{1.5}Gd_{1.5}Fe_{4.5}Al_{0.5}O_{12}$ layer ($4\pi M_s = 360$ **Gs** and $K_U = -1.05 \cdot 10^3$ **erg/cm³**), respectively. The auxiliary layer is needed to deposit high quality main layer with high Bi content and large specific Faraday effect. Each Bragg mirror of the MPC consists of four pairs of alternating 76-nm-thick $TiO_2$ (N1) and 117-nm-thick $SiO_2$ (N2) layers grown on the fused quartz substrate.

**Pump-probe measurements.** To investigate the magnetization precession excitation by laser pulses we used the pump-probe experimental technique. The laser system (Newport Mai Tai HP Ti:Sapphire laser and Spectra-Physics Inspire Auto 100 optical parametric oscillator) emits at 80.54 MHz repetition rate pairs of 150-fs-pulses: the pulse at a wavelength tunable from $\lambda$ = 500 nm to 680 nm and the pulse at $\lambda$ = 820nm. The first one (the pump pulse) is circularly polarized, has light energy fluence of 0.66 mJ/cm$^2$ (calculated for 9 µm diameter) and excites the magnetization precession. The second one (the probe pulse) is 20 times weaker and is used for observation of the magnetization dynamics through the direct Faraday effect, i.e. by measuring variation of the Faraday rotation angle, Ψ, caused by the magnetization precession. The incidence angles of the pump and probe beams are 42º and 30º, respectively. The time delay between the pump and probe pulses is varied from -0.5 to 2.6 ns, where zero time delay corresponds to the simultaneous propagation of the pump and probe pulses through the sample. An external magnetic field, **H**, was applied in-plane to saturate the magnetization.

**Calculation of the electromagnetic field distribution in the MPC.** Distribution of the effective field of the inverse Faraday effect inside the MPC structures (Fig. 3) was calculated by using the transfer matrix method[29]. Dielectric permittivities and magnetic film gyration were taken from[30] and found from fitting the experimentally measured transmittance and Faraday rotation spectra of the MPC with calculated spectra.

At the cavity resonance ($\lambda$ = 642 nm) the magneto-optical parameters of the main and auxiliary layers are $g$=0.0182 and $g$=0.0061, respectively, while the absorption coefficients are $\alpha_1$=2600 cm$^{-1}$ and $\alpha_2$=1990 cm$^{-1}$, respectively. At the wavelength of $\lambda$ = 1400 nm $\alpha_1$=4.9 cm$^{-1}$ and $\alpha_2$=4.0 cm$^{-1}$, and at $\lambda$ = 1841 nm $\alpha_1$=3.9 cm$^{-1}$ and $\alpha_2$=3.2 cm$^{-1}$.

# References

1.	Kirilyuk, A., Kimel, A. V. & Rasing, T. Ultrafast optical manipulation of magnetic order. *Rev. Mod. Phys.* **82,** 2731–2784 (2010).

2.	Beaurepaire, E., Merle, J. C., Daunois, A. & Bigot, J. Y. Ultrafast spin dynamics in ferromagnetic nickel. *Phys. Rev. Lett.* **76,** 4250–4253 (1996).

3.	Lambert, C. H. *et al.* All-optical control of ferromagnetic thin films and nanostructures. *Science* **345,** 1337–1340 (2014).

4.	Bossini, D., Belotelov, V. I., Zvezdin, A. K., Kalish, A. N. & Kimel, A. V. Magnetoplasmonics and Femtosecond Optomagnetism at the Nanoscale. *ACS Photonics* **3,** 1385–1400 (2016).

5.	Kalashnikova, A. M., Kimel, A. V. and Pisarev, R. V. Ultrafast opto-magnetism. *Physics-Uspekhi* **58,** 969 (2015).

6.	Temnov, V. V. Ultrafast acousto-magneto-plasmonics. *Nature Photonics* **6,** 728–736 (2012).

7.	Stupakiewicz, A., Szerenos, K., Afanasiev, D., Kirilyuk, A. & Kimel, A. V. Ultrafast nonthermal photo-magnetic recording in a transparent medium. *Nature* **542,** 71–74 (2017).

8.	Grundler, D. Nanomagnonics. *Journal of Physics D: Applied Physics* **49,** (2016).

9.	Koene, B. *et al.* Spectrally resolved optical probing of laser induced magnetization dynamics in bismuth iron garnet. *J. Phys. Condens. Matter* **28,** 391002 (2016).

10. Deb, M., Vomir, M., Rehspringer, J. L. & Bigot, J. Y. Ultrafast optical control of magnetization dynamics in polycrystalline bismuth doped iron garnet thin films. *Appl. Phys. Lett.* **107,** 252404 (2015).

11. Satoh, T. *et al.* Directional control of spin-wave emission by spatially shaped light. *Nat. Photonics* **6,** 662–666 (2012).

12. Jäckl, M. *et al.* Magnon Accumulation by Clocked Laser Excitation as Source of Long-Range Spin Waves in Transparent Magnetic Films. *Phys. Rev. X* **7,** 21009 (2017).

13. Parchenko, S. *et al.* Non-thermal optical excitation of terahertz-spin precession in a magneto-optical insulator. *Appl. Phys. Lett.* **108,** 32404 (2016).

14. Yoshimine, I., Tanaka, Y. Y., Shimura, T. & Satoh, T. Unidirectional control of optically induced spin waves. *EPL* **117,** 67001 (2017).

15. Chernov, A. I. *et al.* Optical excitation of spin waves in epitaxial iron garnet films: MSSW vs BVMSW. *Opt. Lett.* **42,** 279 (2017).

16. Shelukhin, L. A. *et al.* Ultrafast laser-induced changes of the magnetic anisotropy in a low-symmetry iron garnet film. *Phys. Rev. B* **97,** 14422 (2018).

17. Mikhaylovskiy, R. V., Hendry, E. & Kruglyak, V. V. Ultrafast inverse Faraday effect in a paramagnetic terbium gallium garnet crystal. *Phys. Rev. B - Condens. Matter Mater. Phys.* **86,** 100405(R) (2012).

18. Savochkin, I. V. *et al.* Generation of spin waves by a train of fs-laser pulses: A novel approach for tuning magnon wavelength. *Sci. Rep.* **7,** 5668 (2017).

19. Maccaferri, N. *et al.* Anisotropic Nanoantenna-Based Magnetoplasmonic Crystals for Highly Enhanced and Tunable Magneto-Optical Activity. *Nano Lett.* **16,** 2533–2542 (2016).

20. Kataja, M. *et al.* Surface lattice resonances and magneto-optical response in magnetic nanoparticle arrays. *Nat. Commun.* **6,** 7072 (2015).

21. Armelles, G., Cebollada, A., García-Martín, A. & González, M. U. Magnetoplasmonics: Magnetoplasmonics: Combining Magnetic and Plasmonic Functionalities. *Adv. Opt. Mater.* **1,** 10–35 (2013).

22. Barsukova, M. G. *et al.* Magneto-Optical Response Enhanced by Mie Resonances in Nanoantennas. *ACS Photonics* **4,** 2390–2395 (2017).

23. Ignatyeva, D. O. *et al.* Magneto-optical plasmonic heterostructure with ultranarrow resonance for sensing applications. *Sci. Rep.* **6,** 28077 (2016).

24. Belotelov, V. I. & Zvezdin, A. K. Inverse transverse magneto-optical Kerr effect. *Phys. Rev. B - Condens. Matter Mater. Phys.* **86,** 155133 (2012).

25. Lyubchanskii, I. L., Dadoenkova, N. N., Lyubchanskii, M. I., Shapovalov, E. A. & Rasing, T. Magnetic photonic crystals. *Journal of Physics D: Applied Physics* **36,** R277 (2003).

26. Inoue, M., Fujikawa, R., Baryshev, A., Khanikaev, A., Lim, P. B., Uchida, H., Aktsipetrov, O., Fedyanin, A., Murzina, T., Granovsky A. Magnetophotonic crystals. *J. Phys. D. Appl. Phys.* **39,** R151 (2006).

27. Belotelov, V. I., Kalish, A. N., Kotov, V. A. & Zvezdin, A. K. Slow light phenomenon and extraordinary magnetooptical effects in periodic nanostructured media. *J. Magn. Magn. Mater.* **321,** 826–828 (2009).

28. Liu, T. M. *et al.* Nanoscale Confinement of All-Optical Magnetic Switching in TbFeCo - Competition with Nanoscale Heterogeneity. *Nano Lett.* **15,** 6862–6868 (2015).

29. Le Guyader, L. *et al.* Nanoscale sub-100 picosecond all-optical magnetization switching in GdFeCo microstructures. *Nat. Commun.* **6,** 5839 (2015).

30. Cai, Y. *et al.* Strong enhancement of nano-sized circularly polarized light using an aperture antenna with V-groove structures. *Opt. Lett.* **40,** 1298–1301 (2015).

31. Vasiliev, M. *et al.* RF magnetron sputtered (BiDy)3(FeGa)5O12:Bi2O3 composite garnet-oxide materials possessing record magneto-optic quality in the visible spectral region. *Opt. Express* **17,** 19519–19535 (2009).

32. Prokopov, A. R. *et al.* Epitaxial Bi-Gd-Sc iron-garnet films for magnetophotonic applications. *J. Alloys Compd.* **671,** 403–407 (2016).

33. Kozhaev, M. A. *et al.* Peculiarities of the inverse Faraday effect induced in iron garnet films by femtosecond laser pulses. *JETP Lett.* **104,** 833–837 (2016).

34. Belotelov, V. I. & Zvezdin, A. K. Magneto-optical properties of photonic crystals. *J. Opt. Soc. Am. B* **22,** 286–292 (2005).

35. Yin, Y. *et al.* Tunable permalloy-based films for magnonic devices. *Phys. Rev. B - Condens. Matter Mater. Phys.* **92,** 24427 (2015).

36. Kataja, M. *et al.* Plasmon-induced demagnetization and magnetic switching in nickel nanoparticle arrays. *Appl. Phys. Lett.* **112,** 72406 (2018).

## Acknowledgements

We acknowledge the financial support from Russian Science Foundation (grant No.17-72-20260).

## Author Contribution

The samples were designed and fabricated by A.N.S, A.R.P. and V.N.B., theoretical investigation was performed by D.A.S, A.K.Z. and V.I.B., while M.A.K., A.I.C. and D.A.S performed the measurements and analyzed the data. M.A.K., A.I.C. and V.I.B. wrote the manuscript. All authors discussed the results and contributed to writing of the manuscript.

## Competing interests

The authors declare no competing financial or non-financial interests.